\documentclass[twocolumn, aps, prd, floats, floatfix, amsmath, amssymb, superscriptaddress, preprintnumbers]{revtex4-2}

\usepackage{graphicx}
\usepackage{chemarrow}
\usepackage{mathrsfs}
\usepackage{float}
\usepackage{physics}
\usepackage{latexsym}
\usepackage[colorlinks=true, linkcolor=blue, citecolor=purple]{hyperref}
\usepackage[normalem]{ulem}
\usepackage{orcidlink}
\usepackage{appendix}
\usepackage{braket}

\begin{document}

\preprint{$\begin{gathered}\includegraphics[width=0.05\textwidth]{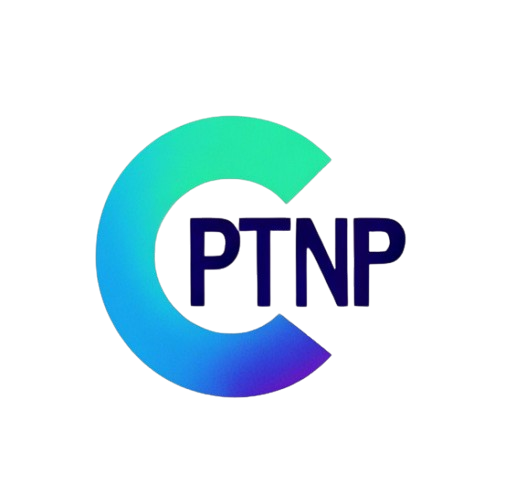}\end{gathered}$\, CPTNP-2025-042}

\title{
Parameter Inference from Final-State Entanglement in Higgs Decays
}

\author{Jia Liu \orcidlink{0000-0001-7386-0253}}
\email{jialiu@pku.edu.cn}
\affiliation{School of Physics and State Key Laboratory of Nuclear Physics and Technology, Peking University, Beijing 100871, China}
\affiliation{Center for High Energy Physics, Peking University, Beijing 100871, China}
	
\author{Masanori Tanaka \orcidlink{0000-0002-1303-7043}}
\email{tanaka@pku.edu.cn}
\affiliation{Center for High Energy Physics, Peking University, Beijing 100871, China}

\author{Xiao-Ping Wang \orcidlink{0000-0002-2258-7741}}
\email{hcwangxiaoping@buaa.edu.cn}
\affiliation{School of Physics, Beihang University, Beijing 100083, China}

\author{Jing-Jun Zhang \orcidlink{0009-0007-5228-5959}}
\email{zhang\_jingjun@stu.pku.edu.cn}
\affiliation{School of Physics and State Key Laboratory of Nuclear Physics and Technology, Peking University, Beijing 100871, China}

\author{Zifan Zheng}
\email{zifanzheng@pku.edu.cn}
\affiliation{School of Physics and State Key Laboratory of Nuclear Physics and Technology, Peking University, Beijing 100871, China}

\begin{abstract}
The decay out-states of unstable Standard Model (SM) particles provide a natural, well-defined quantum-information probe of the SM parameter space. We use Higgs decays as a test case: after tracing out kinematics, we compute entanglement among final-state spins and colors across all decay channels and investigate a near-maximal entanglement-entropy criterion. This criterion turns out to be informative about fundamental parameters. Within the SM, we find that the entanglement entropy exhibits a global maximum close to the observed Higgs mass and the measured $W$ mass, the latter being equivalent to the $SU(2)_L$ gauge coupling. In a two-parameter kappa framework, applying the same criterion is consistent with an SM-like balance between Higgs couplings to vector bosons and to fermions, providing sensitivity to the ratio of the sector-wide rescalings. These results suggest that entanglement extremality may serve as a complementary handle on fundamental parameters.
\end{abstract}

\maketitle

\section{Introduction}

Quantum-information ideas are increasingly applied to particle physics. Since Bell’s proposal of experimentally testable entanglement \cite{Bell:1964kc}, the possibility of probing quantum correlations in fundamental interactions has been widely explored \cite{Barr:2024djo}. Observations of spin entanglement in top–quark pairs at the LHC \cite{ATLAS:2023fsd,CMS:2024pts} have opened a possibility that colliders can probe entanglement directly, establishing quantum-information observables as a viable probe in high-energy physics \cite{Afik:2025ejh,Lykken:2020xtx,Afik:2022kwm}.

Recent studies probe whether quantum-information extrema encode Standard Model (SM) parameters. Examples include maximal helicity entanglement in lepton annihilation/scattering yielding $\sin^2\theta_W=1/4$~\cite{Cervera-Lierta:2017tdt}, minimized entanglement in neutrino oscillations constraining the PMNS CP phase~\cite{Quinta:2022sgq}, and minimized flavor-entanglement power providing insight into CKM and PMNS mixing~\cite{Thaler:2024anb}. Magic–based QED criteria further refine $\theta_W$ inferences \cite{Liu:2025bgw}. Related studies link entanglement to symmetries and dynamics~\cite{Beane:2018oxh, Low:2021ufv, Liu:2022grf, Carena:2023vjc, Liu:2023bnr, Hu:2024hex, Chang:2024wrx, Kowalska:2024kbs, McGinnis:2025brt, Busoni:2025dns, Liu:2025pny, Carena:2025wyh, Hu:2025lua, Nunez:2025xds, Lyu:2025lja, McGinnis:2025iab, McGinnis:2025xgt}.

The Higgs boson underlies electroweak symmetry (EW) breaking and the origin of particle masses. While its properties have been extensively measured at colliders \cite{ATLAS:2022vkf, CMS:2022dwd}, its mass remains a free SM parameter without a universally accepted theoretical explanation. Several approaches aim to explain it: asymptotic-safety arguments in the SM coupled to gravity suggest $m_h \simeq 126$ GeV \cite{Shaposhnikov:2009pv}, criticality of the Higgs potential yields $m_h \sim 135 \pm 9$ GeV \cite{Froggatt:1995rt}, and Planck-scale mass sum rules predict $m_h \sim 135$ GeV\,\cite{Degrassi:2012ry}.
Additionally, maximization of the product of Higgs decay probabilities gives $m_{h} \simeq 125\,{\rm GeV}$\,\cite{dEnterria:2012eip}. An extension of this analysis, using a maximal classical multinomial entropy constructed from the Higgs decay branching ratios, yields $m_{h} = 125.04 \pm 0.25\,{\rm GeV}$~\cite{Alves:2014ksa}.

In this work, we explore how the decays of unstable particles can be informative about SM parameters. Complementary to scattering analyses, which require a specified center-of-mass energy, scattering angle and a two-particle in-state, decays are kinematically fixed in the parent’s rest frame and start from a single-particle in-state determined by the SM spectrum. This makes decay out-states a natural and well-defined quantum-information probe in particle physics. 

We focus on Higgs decays as a test case, analyzing kinematics and evaluating final-state entanglement in spin and color. The Higgs decays provide a natural setting for quantum-information tests, with studies of Bell inequalities and entanglement signatures across multiple channels \cite{Barr:2021zcp, Fabbrichesi:2022ovb, Aguilar-Saavedra:2022wam, Aguilar-Saavedra:2022mpg, Ashby-Pickering:2022umy, Altakach:2022ywa, Fabbrichesi:2023cev, Bernal:2023ruk, Ma:2023yvd, Aguilar-Saavedra:2024whi, Morales:2024jhj, Bernal:2024xhm, Grossi:2024jae, DelGratta:2025qyp, Goncalves:2025qem, Goncalves:2025xer, Bechtle:2025ugc, Abel:2025skj, DelGratta:2025xjp}. By considering all accessible decay channels, rather than a single exclusive mode, we link our observable directly to the Higgs mass, EW gauge couplings, and Yukawa couplings. 
A near-maximal entropy criterion then provides insights into these parameters: within the SM it points to Higgs and $W$ masses close to their observed values. In a two-parameter $\kappa$ framework, it favors an approximately SM-like balance between Higgs couplings to vector bosons and to fermions. This method is broadly applicable, directly tied to measurable decay patterns, and can in principle be applied to other unstable particles beyond the Higgs boson.

\section{The decay out-state of unstable particles}

An unstable particle such as the Higgs boson is prepared in a one-particle in-state. At the rest frame, it is given by
\begin{align}
\ket{\rm in} = \mathcal{N}^{-1} \ket{\mathbf{p}_{\rm in}} \ket{\lambda_{\rm in}} \,, 
\end{align}
where $\ket{\mathbf{p}_{\rm in}}$ denotes a one-particle momentum eigenstate and $\ket{\lambda_{\rm in}}$ collects intrinsic quantum numbers (spin/helicity and gauge charges). 
The normalization factor $\mathcal{N} = (2 E_{\rm in} \mathcal{V})^{1/2}$ ensures $\braket{{\rm in} | {\rm in}} = 1$ with $ \langle \mathbf{p} | \mathbf{p}' \rangle = 2 E_{\rm in} (2\pi)^3 \delta^{(3)}(\mathbf{p}_{\rm in} - \mathbf{p}'_{\rm in}) $ and the box normalization $(2\pi)^3 \delta^{(3)}(\mathbf{0}) = \mathcal{V}$. For a scalar field, the spin label $\lambda_{\rm in}$ is absent.
After the decay, the out-state is
\begin{align}
\ket{\rm out} = i \hat{T} \ket{\rm in} \,, 
\end{align}
with $\hat{T}$ is the transition operator. 
We neglect the trivial identity piece in the $S$-matrix, $S = 1 + i\hat{T}$, since we focus on the entanglement specifically generated \emph{by the decay}.

For two-body decays, the $T$-matrix element is related to the invariant amplitude $\mathcal{M}$ by
\begin{align}
\begin{aligned}
&\braket{\mathbf{p}_{1} \mathbf{p}_{2}; s_{1} s_{2}; c_{1} c_{2}; i j\, | i\hat{T} | \mathbf{p}_{\rm in}, \lambda_{\rm in}} \\ 
& \quad = (2\pi)^4 \delta^{4}(p_{\rm in} - p_{1} - p_{2}) i \mathcal{M}_{s_{1}s_{2}}^{c_{1}c_{2};\, ij}(p_{\rm in} \to p_{1} p_{2}) \,, 
\end{aligned}
\end{align}
where $\ket{s_{\alpha}}$ and $\ket{c_{\alpha}}$ ($\alpha=1,2$) label spin/helicity and color states, and $\ket{i,j}$ labels the particle species in the final state. 
For Higgs decays, the initial particle carries neither color nor spin. 
Color conservation implies $c_{2} = \overline{c_1}$, so we use $\mathcal{M}_{s_{1}s_{2}}^{c_{1}c_{2};\, i,j} \to \mathcal{M}_{s_{1}s_{2}}^{c_{1};\, i,j}$ below.
The out-state can then be written in terms of amplitudes as
\begin{align}
\ket{\rm out}
= ~&\frac{1}{\mathcal{N}} \sum_{i, j}  \sum_{s_{1} s_{2}} \sum_{c} \int d\Pi_{1} d\Pi_{2} (2\pi)^4 \delta^{4}(p_{\rm in} - p_{1} - p_{2}) \nonumber \\
&\times \mathcal{M}_{s_{1} s_{2}}^{c; \, ij}(p_{\rm in} \to p_{1} p_{2}) \ket{\mathbf{p}_{1} \mathbf{p}_{2}; s_{1} s_{2}; c \overline{c}; i j} \,,
\label{eq:out_state}
\end{align}
with $d\Pi_{i} \equiv d^3 \mathbf{p}_{i}/[2E_{i}(2\pi)^3]$. 
From Eq.\,\eqref{eq:out_state}, the normalized density matrix of the final state is
\begin{align}
\label{eq:rho_f}
\rho_{f} = \frac{\ket{\rm out} \bra{\rm out}}{ \braket{{\rm out}|{\rm out}} } \,,
\end{align}
with
\begin{align}
\braket{{\rm out}|{\rm out}}
= \frac{(2\pi)^4 \delta^{4}(0)}{\mathcal{N}^2} 2 m_{\rm in} \Gamma_{\rm tot} = T \Gamma_{\rm tot} \,,
\end{align}
where $\delta(0)$ is the delta function reflecting the plane-wave normalization for momentum eigenstates, $m_{\rm in}$ is the mass of the decaying particle, and $\Gamma_{\rm tot}$ is its total decay width. Here we use the standard box-normalization replacement $(2\pi)\delta(0) \to T$. 

\section{Entanglement Entropy from decays}

We quantify entanglement in the decay out-state using the entanglement entropy (EE), following the formulations in Refs.\,\cite{Lello:2013bva,Seki:2014cgq}. We take the bipartition of final states as $\mathcal{H}_{\rm tot} = \mathcal{H}_{A} \otimes \mathcal{H}_{B}$, with
\begin{align}
& \mathcal{H}_{A} = \mathcal{H}_{\rm spin}^{a} \otimes \mathcal{H}_{\rm color}^{a} \otimes \mathcal{H}_{\rm particle}^{a} \,, \label{eq:HilbertA} \\
&\mathcal{H}_{B} = \mathcal{H}_{\rm kin}^{a} \otimes \mathcal{H}_{\rm kin}^{b} \otimes \mathcal{H}_{\rm spin}^{b} \otimes \mathcal{H}_{\rm color}^{b} \otimes \mathcal{H}_{\rm particle}^{b} \,, 
\label{eq:HilbertB}
\end{align}
where $\mathcal{H}_{\rm kin}^{i}$, $\mathcal{H}_{\rm spin}^{i}$, $\mathcal{H}_{\rm color}^{i}$, and $\mathcal{H}_{\rm particle}^{i} (i=a,b)$ denote the momentum, spin, color, and particle-type Hilbert spaces for particles $i$. 
The reduced density matrix can be got from the total density matrix $\rho$ by tracing out $\mathcal{H}_B$
\begin{align}
\label{eq:reduced_rho}
\rho_{R} = {\rm tr}_{\mathcal{H}_{B}}[\rho] \,, 
\end{align}
and we adopt the linear entropy (Tsallis-2) as our measure of entanglement \cite{Zanardi:2000zz, Low:2024hvn}:
\begin{align}
\label{eq:LinearEE}
EE(\rho_{R})  = 1 - {\rm tr}_{\mathcal{H}_{A}} [\rho_{R}^2] \,. 
\end{align}
Applying $\rho_{f}$ given in Eq.\,\eqref{eq:rho_f} to Eq.\,\eqref{eq:reduced_rho} , the reduced density matrix becomes
\begin{align}
\label{eq:rhoR_2body}
\rho_{R}
=\sum_{i} \sum_{s_{1} s'_{1} s_{2}} \sum_{c}  \frac{\Gamma_{s_{1} s_{2}; s'_{1} s_{2}}^{cc; \, ii}}{\Gamma_{\rm tot}}  \ket{s_{1}, c, i} \bra{s'_{1}, c, i} \,, 
\end{align}
where  
\begin{align}
\Gamma_{s_{1} s_{2} ;s_{3} s_{4}}^{c_{1}c_{2};\,ij}
\equiv \int \frac{d\Phi_{2}(p_{1}, p_{2})}{2m_{\rm in}} \sum_k \mathcal{M}_{s_{1}s_{2}}^{c_{1};\, i k} (\mathcal{M}_{s_{3}s_{4}}^{c_{2}; \, j k})^{*} \,,
\end{align}
with $d\Phi_{2}( p_{1}, p_{2}) \equiv d\Pi_{1} d\Pi_{2} (2\pi)^4 \delta^{4} (p_{\rm in} \to p_{1}, p_{2})$. 
Note $\rho_{R}$ is block diagonal in color and particle type.

Consequently, the linear EE takes the simple form
\begin{align}
EE(\rho_{R}) = 1 - \sum_{i} \frac{\mathcal{P}_{i}}{N_{c}^{i}} {\rm BR}_{i}^2 \,, 
\label{eq:EE}
\end{align}
where $N_{c}^{i}$ and ${\rm BR}_{i}$ are the color multiplicity and the branching ratio (BR) for species $i$. Thus, EE includes a $1/N_c$ color suppression, $1/3$ for quarks and $1/8$ for gluons.
The spin factor $\mathcal{P}_{i}$ is 
\begin{align}
\label{eq:Pi_factor}
\mathcal{P}_{i} = \frac{\sum_{s_{1} s_{2} s_{3} s_4} \Gamma_{s_{1} s_{2} ;s_{3} s_{2}}^{cc; \, ii} \Gamma_{s_{3} s_{4} ; s_{1} s_{4}}^{cc; \, ii}}{ \left(\sum_{s_{1} s_{2}} \Gamma_{s_{1} s_{2} ; s_{1} s_{2}}^{cc; \, ii} \right)^2 } \,. 
\end{align}
For Higgs decays into fermions, this yields $\mathcal{P}_{i}=1/2$.

We emphasize that the EE in Eq.~\eqref{eq:EE} includes contributions from not only the quantum correlation but also the classical correlation. 
The effect of quantum entanglement introduces the additional factors $\mathcal{P}_{i}$ and $N_{c}^{i}$ into the classical entropy formula as shown in Eq.~\eqref{eq:EE}. 
As we discuss later, these weight factors play an important role in reproducing the SM-like EW sector.

\section{Higgs decays in two- and three-body channels}

We apply the decay-based entanglement framework to Higgs decays. 
In the SM, there are 14 Higgs decay channels:
\begin{align}
h \to \, q\bar{q} \,,~ \ell \bar{\ell} \,,~ W W \,,~ Z Z \,, ~ gg \,, ~\gamma \gamma \,,~  Z\gamma \,,
\end{align}
where $q = u \,, d \,, c \,, s \,, t \,, b$ and $\ell = e \,, \mu \,, \tau$~\footnote{
All channels involve the same type particle pair except the $Z\gamma$ mode. The $Z\gamma$ channel would introduce cross-channel interference terms in Eq.~\eqref{eq:EE}. Since ${\rm BR}_{Z\gamma}$ is negligible, we ignore these terms and treat Eq.~\eqref{eq:EE} as exact.}.

When $m_h < 2 m_V$, we use the notation $V^{*}$ indicates an off-shell gauge boson.
The $VV^{*}$ channel is nontrivial for the EE calculation as it yields a three-body final state. 
Consider $h(p_{\rm in})\to V(p_{1})V^{*}(q) \to V(p_{1})f(p_{2})\bar{f}'(p_{3})$, with $q = p_{2} + p_{3}$ and $V = W/Z$. The corresponding out-state is 
\begin{align}
\ket{\rm out}  \ni & ~\mathcal{N}^{-1} \sum_{\{ff'\}} \sum_{\lambda_{1}, s_{2}, s_{2}}  \int d\Phi_{3} \mathcal{M}_{\lambda_{1} s_{2} s_{3}}^{ff',V}(p_{1}, p_{2}, p_{3}) \nonumber \\ 
& \times \ket{\mathbf{p}_{1}, \lambda_{1}}_{V} \ket{\mathbf{p}_{2}, s_{2}, f}  \ket{\mathbf{p}_{3}, s_{3}, f'} \,, 
\end{align}
where $d\Phi_{3} = d\Pi_{1} d\Pi_{2} d\Pi_{3} (2\pi)^4 \delta^{4}(p_{\rm in} - p_{1} - p_{2} - p_{3})$ is the three-body phase space, and $\mathcal{M}_{\lambda_{1} s_{2} s_{3}}^{ff',V}(p_{1}, p_{2}, p_{3})$ is the amplitude, with $\lambda_{1}$ and $s_{2,3}$ parameterizing the helicity states of the on-shell $V$ boson and the fermions~\cite{Keung:1984hn}.

The calculation proceeds analogously to the two-body case. 
We trace over $\mathcal{H}_B$ defined in Eq.~\eqref{eq:HilbertB}, now with the understanding that ``particle $b$'' encompasses both $f$ and $f'$. 
This contributes to the reduced density matrix as
\begin{align}
\label{eq:rhoR_3body}
\rho_{R} \ni  \sum_{\lambda_{1}, \lambda'_{1}} \frac{\Gamma_{\lambda_{1}\lambda'_{1} }^{VV}}{\Gamma_{\rm tot}} \ket{\lambda_{1}}_{V} \bra{\lambda'_1}_{V} \,, \quad (V = W, Z) \,, 
\end{align}
where $\lambda_1, \lambda'_1$ are the helicities of the on-shell $V$, and
\begin{align}
\label{eq:Gamma_3body}
\Gamma_{\lambda_{1}\lambda'_{1}}^{VV}
= \sum_{\{ff'\}} \sum_{s_{2}, s_{3}} \int d \Phi_{3} \mathcal{M}_{\lambda_{1} s_{2} s_{3}}^{ff', \, V} \left(\mathcal{M}_{\lambda'_{1} s_{2} s_{3}}^{ff',\,V}\right)^{*} \,. 
\end{align}
Thus, the reduced density matrix for the process $h \to VV^{*}$ has the same structure as in the two-body case, and the subsequent EE analysis can be carried out similarly.

\begin{figure}[t]
\centering
\includegraphics[width=0.95\linewidth]{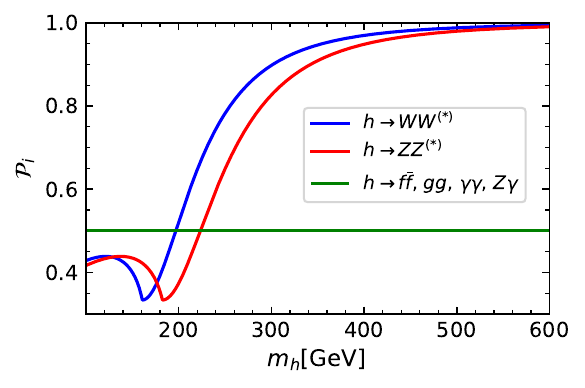}
\caption{
Spin factors $\mathcal{P}_i$ for each Higgs decay channel as a function of $m_h$.
}
\label{fig:Pi_hVV}
\end{figure}

The spin factor $\mathcal{P}_{VV}^{\rm 3-body}$ then follows from Eq.~\eqref{eq:Pi_factor}, see Appendix \hyperref[sec:WWstar-Pi-calculation]{A} for details, 
\begin{align}
\mathcal{P}_{VV}^{\rm 3-body} = \frac{2F_T^2(\epsilon_{V}) + F_L^2(\epsilon_{V})}{[2 F_T(\epsilon_{V}) + F_L(\epsilon_{V})]^2} \,, 
\label{eq:P_VV}
\end{align}
where $F_T$ and $F_L$ encode the transverse and longitudinal helicity contributions,
\begin{align}
F_T(\epsilon_{V})
=& \int_{0}^{y_{\rm max}} dy  \frac{y}{(y-1)^2} \sqrt{ Y - 4 y}  \,,  \\
F_L(\epsilon_{V})
=& \int_{0}^{y_{\rm max}}  dy \frac{1}{4(y-1)^2} Y  \sqrt{ Y - 4 y} \,,  
\end{align}
with $\epsilon_{V} = m_{h}/m_{V} \in [1,\,2]$, $Y = ( \epsilon_{V}^2 - 1 - y )^2 $ and $y_{\rm max} = (\epsilon_{V} - 1)^2$.
Note that $\mathcal{P}_{VV}$ has the same form for the $WW$ and $ZZ$ channels, as the couplings cancel in the ratio.
When $m_h \geq 2 m_V$, the decay $h \to VV$ is two-body and the spin factor is
\begin{align}
\mathcal{P}_{VV}^{\rm 2-body} = \frac{2+r_{L}^2}{(2 + r_{L})^2} \,, 
\end{align}
with $r_{L} = (1 - \epsilon_{V}^2/2)^2$ from the longitudinal contribution. 

The spin factor for each decay channel is shown in Fig.~\ref{fig:Pi_hVV}. 
For $m_h < 2m_V$, $\mathcal{P}_{VV}$ is smaller than $1/2$, which reduces its weight compared to fermion channels. At the threshold $m_h=2m_V$, the two- and three-body descriptions coincide and yield $\mathcal{P}_{VV}=1/3$. Near threshold, the vector bosons are non-relativistic, suppressing helicity information and equalizing contributions from all three helicities. In the heavy-Higgs limit, the vector bosons are highly boosted, and the longitudinal polarization dominates, so
$\mathcal{P}_{VV}\to 1$. 
The channels to $f\bar{f}$, $gg$, $\gamma\gamma$ and $Z\gamma$ have $\mathcal{P}_i=1/2$, since these particles carry two helicities, whereas the longitudinal mode of $Z$ does not contribute. 

\section{Near-maximal EE in Higgs decays}

With the spin factors $\mathcal{P}_i$ determined, we calculate the EE for fixed SM inputs. Higgs BRs are taken from \texttt{HDECAY}~\cite{Djouadi:1997yw, Djouadi:2018xqq}, including only to the three-body modes for $VV^{(*)}$. We also get an independent analytic calculation of partial widths (Appendix~\hyperref[sec:higgs-decay-partial-width]{B}) which gives consistent EE results (Appendix~\hyperref[sec:analytic_EE]{C}).

\begin{figure*}[t]
\centering
\includegraphics[width=0.48\linewidth]{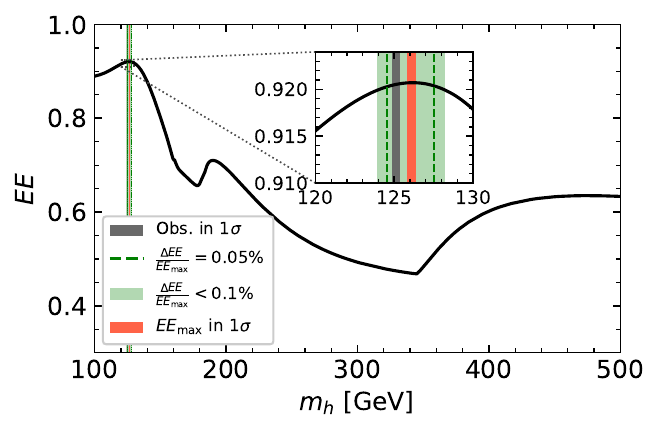} 
\includegraphics[width=0.48\linewidth]{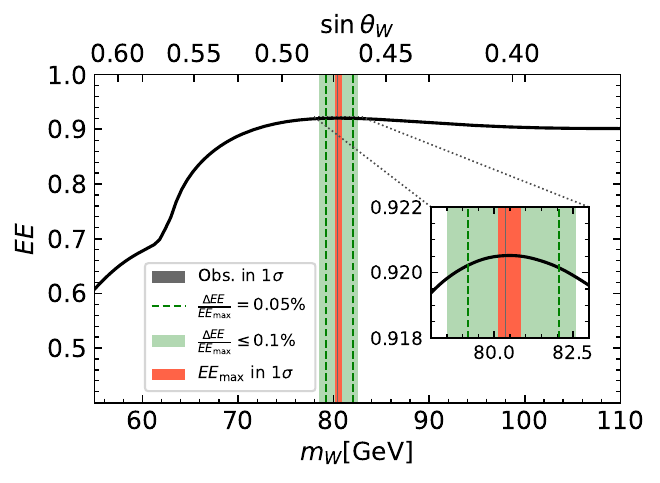} 
\caption{
Entanglement entropy of Higgs decays as a function of the Higgs mass (left) and $m_W$ (right). The gray band marks the experimentally observed SM values ($1\sigma$), the red band shows the $EE_{\max}$ yield with its $1\sigma$ uncertainty from SM inputs and theory, and the green band (dashed lines) indicates the near-maximal-EE region with $\Delta EE/EE_{\max} < 0.1\,\% \, (0.05\,\%)$.
}
\label{fig:Higgs_mass-g}
\end{figure*}

In Fig.~\ref{fig:Higgs_mass-g}, the left panel shows the dependence of the EE on the Higgs mass, with other couplings and the SM vev fixed. A global maximum $EE$ globally gives
\begin{align}
\label{eq:mh_EE}
m_{h}= 126.08 \pm 0.28~{\rm GeV},
\end{align}
where the uncertainty is obtained from a Monte Carlo scan over all other SM parameters within their $1\sigma$ uncertainties~\footnote{
In our analysis, we performed the parameter scan by changing quark masses, lepton masses, strong gauge coupling, $W$ and $Z$ boson masses, Weinberg angle, Fermi constant. 
We used their central values and corresponding ambiguities summarized in Ref.~\cite{ParticleDataGroup:2024cfk} as the input values. 
The theoretical uncertainty in the estimation of BRs is also taken into account~\cite{LHCHiggsCrossSectionWorkingGroup:2016ypw, Bagnaschi:2025lyh}.
For the parameter scan, we assumed that each input parameter takes any value within the uncertainty range with equal probability, and obtained parameter points at which $EE$ is maximized using input values randomly selected within the experimentally allowed range.
The uncertainty of the inferred Higgs and $W$ boson masses is obtained from the central $68\%$ interval of the resulting distribution of the entropy maximum points.
} and the theoretical uncertainty for BRs~\cite{LHCHiggsCrossSectionWorkingGroup:2016ypw, Bagnaschi:2025lyh}. 
This is within $3.2\,\text{-}\,3.3\,\sigma$ of the latest ATLAS result, $125.11 \pm 0.11~{\rm GeV}$~\cite{ATLAS:2023oaq}, and CMS result, $125.08 \pm 0.12~{\rm GeV}$~\cite{CMS:2024eka}. 
Introducing a tolerance for the maximum $EE$ point characterized by
\begin{align}
\frac{\Delta EE}{EE_{\max}} \equiv \frac{|EE(\rho_{R}) - EE_{\max}|}{EE_{\max} } \,, 
\end{align}
the observed Higgs mass lies within this region with $\Delta EE/EE_{\max} \leq 0.05\%$.

The right panel of Fig.~\ref{fig:Higgs_mass-g} shows the EE as a function of $m_W$, with other parameters (including $U(1)_Y$ coupling $g'$) fixed. Equivalently, the EE can be viewed as a function of the $SU(2)_L$ coupling $g$ or the weak mixing angle $\sin\theta_W$. The global maximum EE selects
\begin{align}
\label{eq:mw_EE}
m_{W} = 80.506 \pm 0.376 \, {\rm GeV} \,, 
\end{align}
where the error again reflects the uncertainties in the inputs. The inferred $W$ boson mass value differs from the measured $m_W$ within $0.36\,\sigma$. The corresponding near-maximal-EE band for $m_W$ is considerably broader than input uncertainties.

In Table~\ref{tab:Higgs_W_mass_range}, we summarize the inferred Higgs and $W$ boson mass ranges from the near-maximal EE requirement with each tolerance. 
Since the slight change in the maximal EE point significantly affects the predicted values of $m_{h}$ and $m_{W}$, the difference between our result about $m_{h}$ and its measured value may be explained by considering effects of new physics on Higgs decays.

\begin{table}[t]
    \centering
    \begin{tabular}{|c|c|c|}
    \hline
    $\Delta EE/EE_{\rm max}$ & $m_{h}\,[{\rm GeV}]$ & $m_{W}\,[{\rm GeV}]$ \\
    \hline
    $<0.01\%$ & $(125.51 \,, 126.76)$ & $(79.738 \,, 80.883)$ \\
    $<0.05\%$ & $(124.51 \,, 127.51)$ & $(79.160 \,, 82.041)$ \\
    $<0.1\%$  & $(123.76 \,, 128.26)$ & $(78.581 \,, 82.619)$ \\
    $<0.5\%$  & $(120.51 \,, 130.77)$ & $(76.267 \,, 86.669)$ \\
    \hline
    \end{tabular}
    \caption{
    Inferred Higgs and $W$ boson mass ranges for each tolerance of the maximal EE point.
    }
    \label{tab:Higgs_W_mass_range}
\end{table}

\begin{figure}[t]
\centering
\includegraphics[width=0.95\linewidth]{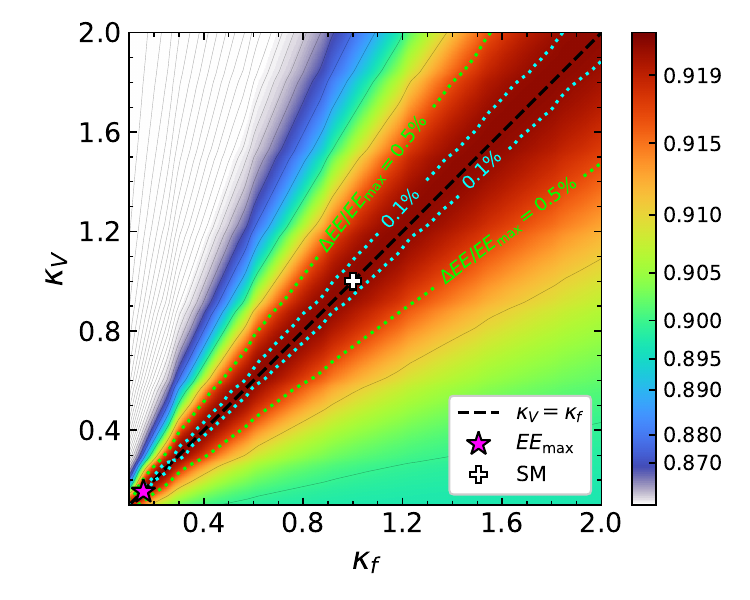} 
\caption{
EE in the $\kappa_f$–$\kappa_V$ plane. Green and cyan dotted curves show contours of $\Delta EE/EE_{\max} = 0.5\%$ and $0.1\%$, respectively. The magenta star marks the maximal-EE point, and the white cross denotes the SM point.
}
\label{fig:2dimplot_kf_kv}
\end{figure}

In Fig.~\ref{fig:2dimplot_kf_kv}, we use the two-parameter $\kappa$ framework~\cite{LHCHiggsCrossSectionWorkingGroup:2013rie} to probe the impact of Higgs-coupling deviations, representative of beyond-SM scenarios such as two-Higgs-doublet models~\cite{Branco:2011iw}. We rescale the SM Higgs couplings as $g_{hVV}\to \kappa_{V} \times g_{hVV}^{\rm SM}$ and $g_{hff}\to \kappa_{f} \times g_{hff}^{\rm SM}$, keeping fermion and vector-boson masses fixed. The figure shows the EE in the $(\kappa_f,\kappa_V)$ plane with all other parameters at their SM values. Assuming a tolerance within $\Delta EE/EE_{\max} \leq 0.1\%$, we obtain
\begin{align}
\frac{\kappa_V}{\kappa_f} = 1.00^{+0.08}_{-0.06} \,.
\end{align}
Thus, the maximal-EE criterion prefers an approximately SM-like balance between $hVV$ and $hff$ couplings. As expected for observables built from BRs, the result is largely insensitive to an overall common rescaling of the couplings.
In other words, our results give an interpretation that new physics models with extended Higgs sectors should satisfy $\kappa_{V} \simeq \kappa_{f}$ to keep the maximal entropy in the SM-like Higgs decays. 
This offers a new insight into which extended Higgs theories are preferred from the perspective of quantum information theory.

\section{Discussions}

We now clarify several conceptual and practical choices in our analysis.
First, as a comparison we also tested an EE constructed from a purely diagonal “branching-ratio density matrix” $\rho_{\rm br} = \sum_i {\rm BR}_i \ket{i}\bra{i}$, with $i$ running over all decay channels. This gives $EE(\rho_{\rm br}) = 1 - \sum_i {\rm BR}_i^2$ and leads to $m_h \sim 135~{\rm GeV}$, an incorrect $m_W \sim 72~{\rm GeV}$, and $\kappa_V/\kappa_f \sim 1.6$, as shown in Fig.~\ref{fig:higgs_mh_1BR} of Appendix~\hyperref[sec:EE_1BR]{D}. These values are different from the observed SM results, indicating that BR-only information is not sufficient and that the spin- and color–derived weights $\mathcal{P}_i/N_c^{i}$ play a crucial role.

Second, as another baseline we evaluate the asymptotic Gibbs-Shannon entropy $S_{\infty} \sim \sum_i\log \mathrm{BR}_i$ of the classical multinomial distribution of Higgs decay counts used in Refs.~\cite{dEnterria:2012eip, Alves:2014ksa}. We crosschecked that it selects $m_{h} \sim 126.52 \pm 0.09~\text{GeV}$ and $m_{W} \sim 81.86 \pm 0.15~\text{GeV}$, close but not fully consistent with the SM value, together with a sizable deviation in $\kappa_V/\kappa_f = 1.3^{+0.2}_{-0.2}$, as shown in Fig.~\ref{fig:g2_mh_Shannon} of Appendix~\hyperref[sec:EE_Shannon]{E}.
Since we include only three-body decays for the $VV$ channels, $m_h$ differs from the $\sim 125~\text{GeV}$ found in Refs.\,\cite{dEnterria:2012eip, Alves:2014ksa}. When the four-body off-shell $VV$ decays are included via \texttt{HDECAY}, we obtain $m_{h} \simeq 125.4\,{\rm GeV}$, which agrees with Ref.\,\cite{Alves:2014ksa}. However, we obtain $m_{W} \simeq 82.9 \pm 0.1\,{\rm GeV}$ and $\kappa_{V}/\kappa_{f} \simeq 1.23$, which deviate from the measured values. For $S_\infty$, the negligible $\text{BR}(h\to e^+e^-)$ contributes to the $\kappa_f$ variation on a similar footing as the dominant $\text{BR}(h\to b\bar b)$, which is counterintuitive: an essentially unobservable channel should not appreciably affect the inferred results.
Since this construction uses only classical branching ratios, we regard it as a complementary cross-check, rather than a full replacement for our entanglement-based approach.

Third, we performed a preliminary study of $W$ and $Z$ decays with unpolarized initial states. For $W$ decays, the $SU(2)_L$ gauge coupling drops out of the EE, leaving only CKM factors and fermion masses; the extremum is essentially flat and does not provide a clear inference. For $Z$ decays, the EE does depend on $\sin\theta_W$, but the preferred value is noticeably shifted from the SM one, as shown in Fig.~\ref{fig:Z_boson_sW} of Appendix~\hyperref[sec:Vboson_decay]{F}. By contrast, Higgs decays appear more informative, since they simultaneously probe the scalar self-coupling, gauge couplings, and Yukawa couplings. Taken together, these observations tentatively suggest that EE in decay processes may be most powerful when it captures \emph{global} information across different sectors, and may be less discriminating within a single sector alone; alternative quantum-information measures beyond EE might be needed there.

Fourth, we deliberately trace out all kinematics at the outset rather than assigning $\mathcal{H}_{\rm kin}^a$ to $\mathcal{H}_A$ for a ``symmetric” bipartition. With plane-wave states, a symmetric split leaves explicit box–normalization factors $\mathcal{V}$ from $\delta^{(3)}(\mathbf{0})$ and $T$ from $\delta(0)$ in the EE. In practice, the volume/time dependence differs between two-body decays ($\propto T/\mathcal{V}$) and three-body decays ($\propto 1/\mathcal{V}$). Thus, the EE diverges or vanishes as $T,\mathcal{V}\to\infty$, and ``normalizing” procedures~\cite{Liu:2025pny} does not cure the mismatch. By tracing out kinematics, the reduced density matrix $\rho_R$ is finite-dimensional (spin, color, species), and the EE is free of $\mathcal{V}$ and $T$. Thus, this bipartition is a well-posed choice. 
Actually, we have numerically checked that this bipartite system can only give a good prediction for the structure of the EW sector.

Finally, we note that full off-shell four-body decays, $h\to V^{*}V^{*}\to 4f$, are not included in our main analysis. Their omission is not due to phenomenological irrelevance, but because a consistent inclusion would require an unambiguous Hilbert-space partition for the four-fermion final state, including the assignment of fermions to the intermediate vector-boson branches, as well as a treatment of identical-fermion and possible $WW/ZZ$ interference effects. In addition, our numerical setup is based on \texttt{HDECAY}, while a dedicated treatment of $h\to 4f$ decays would require a tool such as \texttt{PROPHECY4F}~\cite{Denner:2019fcr}. We therefore treat these modes as a limitation of the present analysis and estimate their possible quantitative impact in Appendix~\hyperref[sec:EE_Shannon]{E}.

\section{Conclusions}

We explore the decay out-states of unstable particles as a natural quantum-information probe of the SM parameter space, formulating an entanglement-entropy quantity and applying it to SM Higgs decays. With spin– and color–entanglement weights included, a near-maximal-EE criterion selects a Higgs mass and $W$ mass close to their measured values, and in a two-parameter $\kappa$ framework it favors an approximately SM-like balance $\kappa_V/\kappa_f \sim 1$. Comparisons with purely branching-ratio–based constructions and with multinomial Shannon entropy show that entanglement weights play a crucial role in obtaining phenomenologically viable inferences. Preliminary $W/Z$ studies further suggest that decay EE may be most informative when it probes \emph{global} scalar–gauge–Yukawa structure, and may be less discriminating within a single sector alone. 
Taken together, these results indicate that the entanglement extremality in decay out-states can act as a compact, data-adjacent probe of SM couplings, and they motivate exploring complementary quantum-information measures beyond the EE when extending this program to other unstable particles.

\begin{acknowledgments}
The authors would like to thank Yuxin He and Jian Wang for helpful discussions.
The work of J.L. is supported by the National Science Foundation of China under Grant No. 12235001, No. 12475103 and State Key Laboratory of Nuclear Physics and Technology under Grant No. NPT2025ZX11.
The work of X.P.W. is supported by the National Science Foundation of China under Grant No. 12375095, and the Fundamental Research Funds for the Central Universities. The work of J.J.Z. is supported by the National Science Foundation of China under Grants No. 11635001, 11875072. J.L. and X.P.W. also thank APCTP, Pohang, Korea, for their hospitality during the focus program [APCTP-2025-F01], from which this work greatly benefited. The authors gratefully acknowledge the valuable discussions and insights provided by the members of the Collaboration of Precision Testing and New Physics.
\end{acknowledgments}

\onecolumngrid
\appendix

\section{The spin factor in the off-shell decay $h \to VV^*$ } \label{sec:WWstar-Pi-calculation}

We here describe the detail of the derivation of Eq.\,\eqref{eq:rhoR_3body}. 
For the process $h(p_{\rm in})\to W(p_{1})W^{*}(q) \to V(p_{1})f(p_{2})\bar{f}(p_{3})$, the final state can be written as
\begin{align}
\left. \ket{\rm out} \right|_{h \to WW^{*}}
= \sum_{f} \sum_{\lambda_{1}, \lambda_{2}, s_{2}, s_{3}}  \int d\Phi_{3}(p_{\rm in}, p_{1}, p_{2}, p_{3})\mathcal{M}_{\lambda_{1} \lambda_{2}s_{2} s_{3}}^{f, W}(p_{1}, p_{2}, p_{3}) \ket{\mathbf{p}_{1}, \lambda_{1}}_{V} \ket{\mathbf{p}_{2}, s_{2}, f}  \ket{\mathbf{p}_{3}, s_{3}, f} \,, 
\end{align}
where the amplitude is given by~\cite{Keung:1984hn}
\begin{align}
\mathcal{M}_{\lambda_{1}\lambda_{2} s_{2} s_{3}}^{f, W} (p_{1}, p_{2}, p_{3}) 
&= \mathcal{M}^{h\to WW^{*}}_{\lambda_{1} \lambda_{2}}(p_{1}, q)  \frac{1}{q^2 - m_{W}^2 + i m_{W} \Gamma_{W}} \mathcal{M}^{W^{*} \to f\bar{f}}_{\lambda_{2} s_{2} s_{3}} (p_{2}, p_{3}) \,, \\ 
\mathcal{M}^{h\to WW^{*}}_{\lambda_{1} \lambda_{2} }(p_{1}, q) 
&= i \frac{2m_{W}^2}{v} g^{\mu \nu} \epsilon_{\mu}^{*}(p_{1}, \lambda_{1}) \epsilon_{\nu}^{*}(q, \lambda_{2})  \,, \\ 
\mathcal{M}^{W^{*} \to f\bar{f}}_{\lambda_{2} s_{2} s_{3}} (p_{2}, p_{3}) 
&= -i g_{2} \epsilon_{\rho}(q, \lambda_{2})\bar{u}_{f}(p_{2}, s_{2}) \gamma^{\rho} \left( \frac{1 - \gamma_{5}}{2} \right) v_{f}(p_{3}, s_{3}) \,,
\end{align}
where $\epsilon(p, \lambda)$ is the polarization vector, and $u_{f}$ and $v_{f}$ are the fermion spinors. 
Although we explicitly keep the helicity state of the off-shell $W^{*}$ boson labeled by $\lambda_{2}$, the expression can be simplified by using the well-known replacement $\sum_{\lambda_{2}} \epsilon_{\rho}(q, \lambda_{2}) \epsilon_{\nu}^{*}(q, \lambda_{2}) \to - g_{\rho \nu} + q_{\rho} q_{\nu}/q^2$. 
Using this replacement, the amplitude simplifies as $\sum_{\lambda_{2}} \mathcal{M}_{\lambda_{1}\lambda_{2} s_{2} s_{3}}^{f, W} \to \mathcal{M}_{\lambda_{1} s_{2} s_{3}}^{f, W}$, as shown in Eq.\,\eqref{eq:rhoR_3body}. 

In deriving the reduced density matrix given in Eq.\,\eqref{eq:reduced_rho}, we trace over the full fermionic final states as
\begin{align}
\rho_{R} \ni
\sum_{\tilde{f}} \sum_{s_{1}', s_{2}'} \int d \Pi_{1} d \Pi_{2} d\Pi_{3} \bra{\mathbf{p}_{1}}_{W} \bra{\mathbf{p}_{2}, \mathbf{p}_{3}; s_{1}', s_{2}', \tilde{f}} \rho_{f} \ket{\mathbf{p}_{1}}_{W} \ket{\mathbf{p}_{2}, \mathbf{p}_{3}; s_{1}', s_{2}', \tilde{f}} \,. 
\label{eq:rhoR_hVV}
\end{align}
Performing the summations and integrals in Eq.\,\eqref{eq:rhoR_hVV}, we obtain 
\begin{align}
\rho_{R} \ni
\frac{1}{2m_{h} \Gamma_{\rm tot}} \sum_{\lambda_{1}, \lambda_{1}' } \sum_{s_{1}, s_{2}} \int d\Phi_{3}(p_{\rm in}, p_{1}, p_{2}, p_{3}) \mathcal{M}_{\lambda_{1} s_{1} s_{2} }^{f, W} (\mathcal{M}_{\lambda_{1}' s_{1} s_{2} }^{f, W})^{*} \ket{\lambda_{1}}_{W} \bra{\lambda_{1}'}_{W}  =  \sum_{\lambda_{1}, \lambda_{1}'}\frac{\Gamma_{\lambda_{1} \lambda_{1}'}^{WW}}{ \Gamma_{\rm tot}} \ket{\lambda_{1}}_{W} \bra{\lambda_{1}'}_{W}  \,, 
\end{align}
which is the same as Eq.\,\eqref{eq:rhoR_3body} with $V = W$. 
An analogous calculation for the $h \to ZZ^{*}$ decay process leads to the same conclusion.

We then derive the expression of the spin factor $\mathcal{P}_{i}$ for the three-body decay process $h \to V f \bar{f}$. 
For this decay process, $\mathcal{P}_{i}$ is given by 
\begin{align}
\mathcal{P}_{i} = \frac{ \sum_{\lambda_{1}, \lambda_{2}} \Gamma_{\lambda_{1} \lambda_{2}}^{VV} \Gamma_{\lambda_{2} \lambda_{1}}^{VV} }{ \left( \sum_{\lambda_{1}} \Gamma_{\lambda_{1} \lambda_{1}}^{VV} \right)^2} \,. 
\end{align}
Therefore, we need to evaluate the decay width $\Gamma_{\lambda_{1} \lambda_{2}}^{VV}$ for each helicity state of the $V$ boson. 
In the following, we derive the explicit expression of $\Gamma_{\lambda_{1} \lambda_{2}}^{VV}$. 

For the three-body phase-space measure, we can use the decomposition
\begin{align}
d\Phi_{3}(p_{\rm in}, p_{1}, p_{2}, p_{3}) = \frac{dx}{2\pi} d\Phi_{2}(p_{\rm in} \to p_{1}, q) d\Phi_{2} (q \to p_{2}, p_{3}) \,, 
\end{align}
with $x = q^2$.
Then, the decay width $\Gamma_{\lambda_{1} \lambda_{2}}^{VV}$ can be expressed as
\begin{align}
\Gamma_{\lambda_{1} \lambda_{3}}^{VV}
= &\frac{1}{2m_{h}}\sum_{f} \sum_{\lambda_{2}, \lambda'_{2}, s_{2}, s_{3}} \int \frac{dx}{2\pi} d \Phi_{2} (p_{1}, q) \frac{ \mathcal{M}_{\lambda_{1}\lambda_{2}}^{h\to VV^{*}} (\mathcal{M}_{\lambda_{3}\lambda'_{2}}^{h\to VV^{*}})^{*} }{(q^2 - m_{V}^2)^2 + m_{V}^2 \Gamma_{V}^2} \int d\Phi_{2}(q \to p_{2}, p_{3}) \mathcal{M}_{\lambda_{2}s_{2}s_{3}}^{V^{*} \to f \bar{f}} (\mathcal{M}_{\lambda'_{2}s_{2}s_{3}}^{V^{*} \to f \bar{f}})^{*} \\
= & \frac{1}{2m_{h}}\sum_{f} \int \frac{dx}{2\pi} d\Phi_{2} \frac{ k_{V} \,  m_{V}^2 }{(q^2 - m_{V}^2)^2 + m_{V}^2 \Gamma_{V}^2} \epsilon^{*}_{\mu}(\lambda_{1}) \epsilon_{\nu}(\lambda_{3}) \left( g^{\mu \rho} - \frac{q^{\mu} q^{\rho}}{q^2} \right) \left( g^{\nu \sigma} - \frac{q^{\nu} q^{\sigma}}{q^2} \right) \int d\Phi_{2} J_{\rho}^{f} J_{\sigma}^{f} \,, 
\end{align}
where $J_{\mu}^{f} = \bar{u}_{f} (c_{V}^{f} - c_{A}^{f} \gamma^{5}) v_{f}$, and $k_{V}$ is a constant which depends on the type of $V$ boson. 
For the last integral, we can use the following useful relation 
\begin{align}
\sum_{f} \int d\Phi_2(q \to p_{2}, p_{3}) J_{\rho}^{f} J_{\sigma}^{f} = x A_{T}^{V} \left( - g_{\rho \sigma} + \frac{q_{\rho} q_{\sigma}}{q^2} \right) \,, 
\end{align}
with 
\begin{align}
A_{T}^{W} \simeq \sum_{f} \frac{N_{c}^{f}}{12\pi} \,, \quad A_{T}^{Z} = \sum_{f} \frac{N_{c}^{f} \left[ (c_{V}^{f})^2 + (c_{A}^{f})^2 \right]}{12\pi} \,.
\end{align}
As a result, we obtain 
\begin{align}
\Gamma_{\lambda_{1} \lambda_{2}}^{VV}
=&  \frac{1}{2m_{h}} \int \frac{dx}{2\pi} \int d\Phi_{2}(p_{\rm in} \to p_{1}, q) \frac{ k_{V} m_{V}^2 }{(q^2 - m_{V}^2)^2 + m_{V}^2 \Gamma_{V}^2} x A_{T}^{V} \left[ - \epsilon^{*}(\lambda_{1}) \cdot \epsilon(\lambda_{2}) + \frac{ (\epsilon^{*}(\lambda_{1}) \cdot q) (\epsilon(\lambda_{2}) \cdot q) }{q^2}  \right] \nonumber \\ 
=& \frac{1}{2m_{h}}\int \frac{dx}{2\pi} \frac{ k_{V} m_{V}^2 }{(x - m_{V}^2)^2 + m_{V}^2 \Gamma_{V}^2}x A_{T}^{V} \frac{\lambda^{1/2}(m_{h}^2, m_{W}^2, x)}{16 \pi m_{h}^2} \Xi_{\lambda_{1} \lambda_{2}} \,,
\label{eq:Gamma_l1l2}
\end{align}
where 
\begin{align}
\lambda(a,b,c) = a^2 + b^2 + c^2 - 2ab - 2bc - 2ca \,. 
\end{align}
The quantity $\Xi_{\lambda_{1} \lambda_{2}}$ satisfies the following equations for each helicity state
\begin{align}
\Xi_{\pm\pm} = 1 \,,\quad \Xi_{00} = r_{V}(x) = \frac{(m_{h}^2 - m_{V}^2 - x)^2}{4m_{V}^2 x} \,, \quad \Xi_{ij} = 0 ~~ (i \neq j) \,. 
\end{align}
For simplicity, we ignore the decay width in the denominator of Eq.\,\eqref{eq:Gamma_l1l2}. 
This assumption is reasonable in the SM because $m_{V} \gg \Gamma_{V}$~\cite{ParticleDataGroup:2024cfk}. 
Then, the spin factor $\mathcal{P}_{i}$ for the $h \to VV^{*}$ can be expressed as a function of the $V$ boson mass and the Higgs boson mass:
\begin{align}
\mathcal{P}_{VV}^{\rm 3-body} = \frac{2F_{T}^2(\epsilon_{V}) + F_{L}^2(\epsilon_{V})}{[2F_{T}(\epsilon_{V}) + F_{L}(\epsilon_{V})]^2} \,, 
\end{align}
where 
\begin{align}
F_{T}(\epsilon_{V})
&= \int_{0}^{f_{\rm max}} df \frac{f}{(f-1)^2} \sqrt{ ( \epsilon_{V}^2 - 1 - f )^2 - 4f}  \,,  \\
F_{L}(\epsilon_{V})
&= \int_{0}^{f_{\rm max}} df \frac{1}{4(f-1)^2} ( \epsilon_{V}^2 - 1 - f )^2\sqrt{ ( \epsilon_{V}^2 - 1 - f )^2 - 4f} \,, 
\end{align}
with $\epsilon_{V} = m_{h}/m_{V} > 1$ and $f_{\rm max} = (\epsilon_{V} - 1)^2$. 
Using the above result, we can confirm that $\lim_{\epsilon_{V} \to 2}\mathcal{P}_{VV}^{\rm 3-body} \to 1/3$, which corresponds to the threshold limit. 

When the Higgs mass satisfies $m_{h} > 2m_{V}$, the on-shell decay process $h(p_{\rm in})\to V(p_{1})V(p_{2})$ is allowed. 
In this case, the decay width is given by 
\begin{align}
\Gamma_{\lambda_{1} \lambda_{2}}^{VV} 
&= \frac{2m_{V}^4}{m_{h}v^2} \int d\Phi_{2}(p_{_{\rm in}}, p_{1}, p_{2}) \left[ - \epsilon^{*}(\lambda_{1}) \cdot \epsilon(\lambda_{2}) + \frac{ (\epsilon^{*}(\lambda_{1}) \cdot p_{2}) (\epsilon(\lambda_{2}) \cdot p_{2}) }{p_{2}^2}  \right] \nonumber \\
&= \frac{m_{V}^4}{8 \pi m_{h}^3 v^2} \lambda^{1/2}(m_{h}^2, m_{V}^2, m_{V}^2) \Xi_{\lambda_{1} \lambda_{2}}
\end{align}
Therefore, the spin factor is given by
\begin{align}
\mathcal{P}_{VV}^{\rm 2-body} = \left. \frac{ 2 \Xi_{++}^2 + \Xi_{00}^2}{(2 \Xi_{++} + \Xi_{00})^2} \right|_{x=m_{V}^2} = \frac{2+r_{V}^2(m_{V}^2)}{[2 + r_{V}(m_{V}^2)]^2} \,. 
\end{align}
Since $r_{V}(m_{V}^2) = 1$ in the case with $m_{h} = 2m_{V}$, we obtain $\mathcal{P}_{VV}^{\rm 2-body} = \mathcal{P}_{VV}^{\rm 3-body} = 1/3$ at the threshold point.

\section{The analytical formulae for Higgs decay partial widths}
\label{sec:higgs-decay-partial-width}

The following formulae for Higgs decay widths are based on Ref.\,\cite{Djouadi:2005gi} (for recent progress on higher-order corrections, see, see Ref.\,\cite{Spira:2016ztx,Bagnaschi:2025lyh}). 

\begin{itemize}

\item $h \to q \bar{q}$ except for top quarks~\cite{Djouadi:2005gi}
\begin{align}
\Gamma(h \to q \bar{q}) = \frac{3 m_{h}}{8 \pi v^2} \bar{m}_{q}^2(m_{h}) \left[ 1 + \Delta_{qq} + \Delta_{h}^2 \right] \,, 
\end{align}
with 
\begin{align}
&\Delta_{qq} = 5.67 \frac{\bar{\alpha}_{S}}{\pi} + (35.94 - 1.36 N_{f}) \left( \frac{\bar{\alpha}_{S}}{\pi} \right)^2 + (164.17 - 25.77 N_{f} + 0.26 N_{f}^2) \left( \frac{\bar{\alpha}_{S}}{\pi} \right)^3 \,,  \\
&\Delta_{h}^2 = \left( \frac{\bar{\alpha}_{S}}{\pi} \right)^2 \left[ 1.57 - \frac{2}{3} \log \frac{m_{h}^2}{m_{t}^2} + \frac{1}{9} \log^2 \left( \frac{\bar{m}_{q}^2(m_{h})}{m_{h}^2} \right) \right] \,, 
\end{align}
where $\bar{m}_{q}(\mu)$, $m_{t}$, $N_{f}$, and $\bar{\alpha}_{S} = \alpha_{S}(m_{h})$ are the $\overline{MS}$ quark mass of flavor $q$ at the scale $\mu$, the top-quark mass, quark generation numbers and the strong coupling constant at $m_{h}$, respectively. 
The state-of-the-art QCD corrections at $\alpha_s^3$, including full quark-mass dependence, have recently been computed~\cite{Wang:2024ilc}.
For the bottom quark, its mass at the Higgs mass scale has been measured at the LHC with a relatively large uncertainty~\cite{Aparisi:2021tym}. 
In our analysis, we employ the renormalization group evolution for quark masses discussed in Ref.\,\cite{Vermaseren:1997fq} with the following input values~\cite{ParticleDataGroup:2024cfk}
\begin{align}
\label{eq:mb_mc_ms}
\bar{m}_{b}(\bar{m}_{b}) = 4.18 \pm 0.03\, {\rm GeV} \,, ~~ 
\bar{m}_{c}(\bar{m}_{c}) = 1.280 \pm 0.025 \,{\rm GeV} \,, ~~
\bar{m}_{s}(2\,{\rm GeV}) = 93.5 \pm 0.8 \,{\rm MeV} \,. 
\end{align}
For the renormalization group evolution of $\alpha_{S}$, it is described at the three-loop level by~\cite{Larin:1993tp, vanRitbergen:1997va}
\begin{align}
\label{eq:alphaS_RG}
\frac{\partial \alpha_{S}}{\partial \ln \mu^2} = - \beta_{0} \alpha_{S}^2 - \beta_{1} \alpha_{S}^3 - \beta_{2} \alpha_{S}^4 \,, 
\end{align}
with 
\begin{align}
\beta_{0} = 11 - \frac{2}{3} N_{f} \,, \quad \beta_{1} = 102 - \frac{38}{3} N_{f} \,, \quad \beta_{2} = \frac{2857}{2} - \frac{5033}{18} N_{f} + \frac{325}{54} N_{f}^2 \,. 
\end{align}
For the input value in Eq.\,\eqref{eq:alphaS_RG}, we use $\alpha_{S}(m_{Z}) = 0.1180 \pm 0.0009$~\cite{ParticleDataGroup:2024cfk}.

\item $h \to t \bar{t}$ \cite{Djouadi:2005gi}
\begin{align}
\Gamma( h \to t \bar{t}) = \frac{3 m_{h}}{8 \pi v^2} m_{t}^2 \beta_{t}^{3} \left[ 1 + \frac{4}{3} \frac{\alpha_{S}}{\pi} \Delta_{H}^{t}(\beta_{t}) \right] \,, 
\end{align}
where $\beta_{i} = \sqrt{ 1 - 4m_{i}^2/m_{h}^2}$, and 
\begin{align}
\Delta_{H}^{t}(\beta)  = \frac{1}{\beta} A(\beta) + \frac{1}{16\beta^3} (3 + 34 \beta^2 - 13\beta^4) \log \left( \frac{1 + \beta}{1 - \beta} \right) + \frac{3}{8 \beta^2} (7\beta^2 - 1) \,. 
\end{align}
The function $A(\beta)$ is defined by 
\begin{align}
\begin{aligned}
A(\beta)  = \, & (1 + \beta^2) \left[ 4 {\rm Li}_{2}\left( \frac{1 - \beta}{1 + \beta} \right) + 2 {\rm Li}_{2} \left( - \frac{1 - \beta}{1 + \beta} \right) - 3 \log \left( \frac{1 + \beta}{1 - \beta} \right) \log \left( \frac{2}{1 + \beta} \right) - 2 \log \left( \frac{1 + \beta}{1 - \beta} \right) \log \beta \right]  \\
& - 3 \beta \log \left( \frac{4}{1 - \beta^2} \right) - 4 \beta \log \beta \,. 
\end{aligned}
\end{align}
where ${\rm Li}_{2}$ is the Spence function.

\item $h \to \ell \bar{\ell}$ \cite{Ellis:1975ap}
\begin{align}
\Gamma( h \to \ell \bar{\ell}) = \frac{m_{h}}{8 \pi v^2} m_{\ell}^2 \beta_{\ell}^{3} \,. 
\end{align}

\item $h \to WW$ (On-shell) \cite{Lee:1977eg} 
\begin{align}
\Gamma(h \to WW) = \frac{g_{2}^2 m_{h}^3}{64\pi m_{W}^2} \sqrt{1 - \frac{4m_{W}^2}{m_{h}^2}} \left( 1 - \frac{4m_{W}^2}{m_{h}^2} + 12 \frac{m_{W}^4}{m_{h}^4} \right) \,. 
\end{align}
Radiative corrections in this process require inclusion of the soft-photon bremsstrahlung contribution to avoid infrared divergences~\cite{Kniehl:1991xe}. 
Since tracing out the soft-photon states is non-trivial, we employ the above formula in our analysis in the region $m_{h}>2m_{W}$. 

\item $h \to WW^{*}$ (Off-shell) \cite{Keung:1984hn}
\begin{align}
\Gamma(h \to WW^{*}) 
= \frac{3 g_{2}^4 m_{h}}{512 \pi^3} F\left( \frac{m_{W}}{m_{h}} \right) \,, 
\end{align}
with
\begin{align}
\begin{aligned}
F(x) = \frac{3(1-8x^2 + 20x^4)}{\sqrt{4x^2 - 1}} \arccos \left( \frac{3x^2 - 1}{2x^3} \right)
- (1 - x^2) \left( \frac{47}{2} x^2 - \frac{13}{2} + \frac{1}{x^2} \right) 
- \frac{3}{2}(1 - 6x^2 + 4x^4) \ln x^2 \,.
\end{aligned}
\end{align}

In addition to the decay process $h \to VV^{*}$, the Higgs boson can also decay via two off-shell gauge bosons, $h \to V^{*} V^{*}$~\cite{Bredenstein:2006rh}. These processes can modify ${\rm BR}(h \to VV)$ by a few percent~\cite{Djouadi:2005gi, Bredenstein:2006rh}. This effect is neglected in our analyses since it is numerically subdominant, \texttt{HDECAY} does not include the full $WW/ZZ$ and identical-fermion interference, and our $\mathcal{P}_i$ construction would mix the $WW$ and $ZZ$ channels in $\rho_R$ and introduce an ambiguous choice of fermion partition for four-body final states.

In addition, we treat the final–state fermions as massless. This is an excellent approximation in the SM: helicity–flip terms in the $V^{*}\to f\bar f'$ current are proportional to $m_f$, so both rate and polarization effects enter only at $O(m_f^{2}/m_V^{2})$. For the bottom quark case, we obtain $(m_b/m_W)^{2}\simeq 2.7\times10^{-3}$ with all other fermions much smaller, and phase–space corrections from finite $m_f$ are of similar size. Since these effects are well below the quoted input-parameter uncertainties, $\Delta m_b/\bar{m}_b(\bar{m}_{b}) \sim 0.7\%$, the massless approximation is justified.

\item $h \to ZZ$ (On-shell) \cite{Lee:1977eg}
\begin{align}
\Gamma(h \to ZZ) = \frac{g_{2}^2 m_{h}^3}{128 \pi m_{Z}^2} \sqrt{1 - \frac{4m_{Z}^2}{m_{h}^2}} \left( 1 - \frac{4m_{Z}^2}{m_{h}^2} + 12 \frac{m_{Z}^4}{m_{h}^4} \right) \,. 
\end{align}

\item $h \to ZZ^{*}$ (Off-shell) \cite{Keung:1984hn}
\begin{align}
\Gamma(h \to ZZ^{*}) 
= \frac{g_{2}^4 m_{h}}{2048 \pi^3 \cos \theta_{W}^4} \left( 7 - \frac{40}{3} \sin \theta_{W}^2 + \frac{160}{9} \sin \theta_{W}^4 \right) F\left( \frac{m_{Z}}{m_{h}} \right) \,, 
\end{align}

\item $h \to \gamma \gamma$ \cite{Ellis:1975ap, Shifman:1979eb}
\begin{align}
\Gamma(h \to \gamma \gamma) 
= \frac{\alpha_{\rm EM}^2}{256 \pi^3 v^2} m_{h}^3 \left| \sum_{f} N_{c}^{f} Q_{f}^2 A_{1/2}(\tau_{f}) + A_{1}(\tau_{W}) \right|^2 \,,
\end{align}
where $N_{c}^{f}$ is the color degrees of freedom for the fermion $f$, and 
\begin{align}
A_{1}(x) = 2 + 3 x + 3x(2-x) f(x) \,, \quad A_{1/2}(x) = -2 x \left[ 1 + (1-x) f(x) \right] \,, 
\end{align}
with $\tau_{X} = 4m_{X}^2/m_{h}^2$, and 
\begin{align}
f(x) = \begin{cases}
\arcsin^2 (x^{-1/2}) \quad &(x \leq 1) \\
-\frac{1}{4} \left[ \log \left( \frac{1+\sqrt{1-x}}{1-\sqrt{1-x}} \right) - i \pi \right]^2 \quad &(x > 1)
\end{cases} \,. 
\end{align}

\item $h \to Z \gamma$ \cite{Cahn:1978nz}
\begin{align}
\Gamma(h \to Z \gamma) 
= \frac{m_{W}^2 \alpha_{\rm EM}^{} m_{h}^3}{128 v^4 \pi^4} \left(1 - \frac{m_{Z}^2}{m_{h}}\right)^3 \left| \sum_{f} N_{f}^{c} \frac{Q_{f} \hat{v}_{f}}{c_{W}} A_{1/2}(\tau_{f}, \lambda_{f}) + A_{1}(\tau_{W}, \lambda_{W}) \right|^2 
\end{align}
where $\lambda_{i} \equiv 4 m_{i}^2/m_{Z}^2$, $\hat{v}_{f} = 2 T_{f}^{3} - 4 Q_{f} s_{W}^2$, and 
\begin{align}
&A_{1/2}(x, y) = I_{1}(x, y) - I_{2}(x, y) \,, \\
&A_{1}(x, y) = c_{W} \left[ 4 \left( 3 - \frac{s_{W}^2}{c_{W}^2} \right) I_{2}(x,y) + \left\{ \left( 1 + \frac{2}{x} \right) \frac{s_{W}^2}{c_{W}^2} - \left( 5 + \frac{2}{x} \right) \right\} I_{1}(x, y)\right] \,, 
\end{align}
with 
\begin{align}
&I_{1}(x, y) = \frac{x y}{2(x -y)} + \frac{x^2 y^2}{2(x - y)^2} \left[ f(x^{-1}) - f(y^{-1}) \right] + \frac{x^2 y}{(x-y)^2} \left[ g(x^{-1}) - g(y^{-1}) \right] \,, \\
&I_{2}(x, y) = -\frac{x y}{2(x - y)} \left[ f(x^{-1}) - f(y^{-1}) \right] \,, \\
&g(x) = \begin{cases}
\sqrt{x^{-1} -1} \arcsin \sqrt{x} \quad &(x \geq 1) \\ 
\frac{ \sqrt{1 - x^{-1}}}{2} \left[ \log \left( \frac{1+\sqrt{1-x^{-1}}}{1-\sqrt{1-x^{-1}}} \right) - i \pi \right] \quad &(x < 1)
\end{cases} \,. 
\end{align}

\item $h \to gg$ \cite{Rizzo:1979mf}
\begin{align}
\Gamma( h \to gg) = \frac{\alpha_{S}^2 m_{h}^3}{72 v^2 \pi^3} \left| \frac{3}{4} \sum_{q} A_{1/2}(\tau_{q}) \right|^2 \,. 
\end{align}

\end{itemize}

\section{Numerical results from analytic decay widths }
\label{sec:analytic_EE}

To check consistency, we compare the EE obtained from \texttt{HDECAY} with that from the analytic expressions for the Higgs decay widths summarized in Appendix~\hyperref[sec:higgs-decay-partial-width]{B}.

\begin{figure}
\centering
\includegraphics[width=0.48\linewidth]{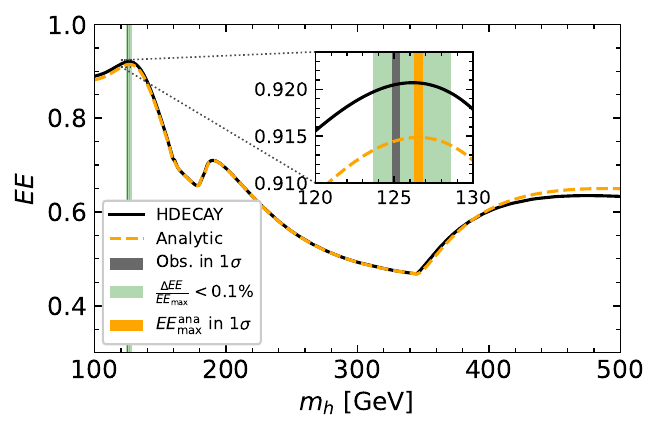}
\includegraphics[width=0.48\linewidth]{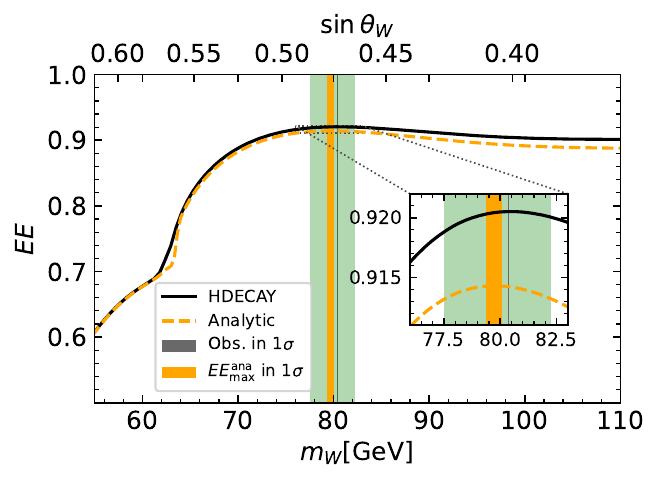}
\includegraphics[width=0.48\linewidth]{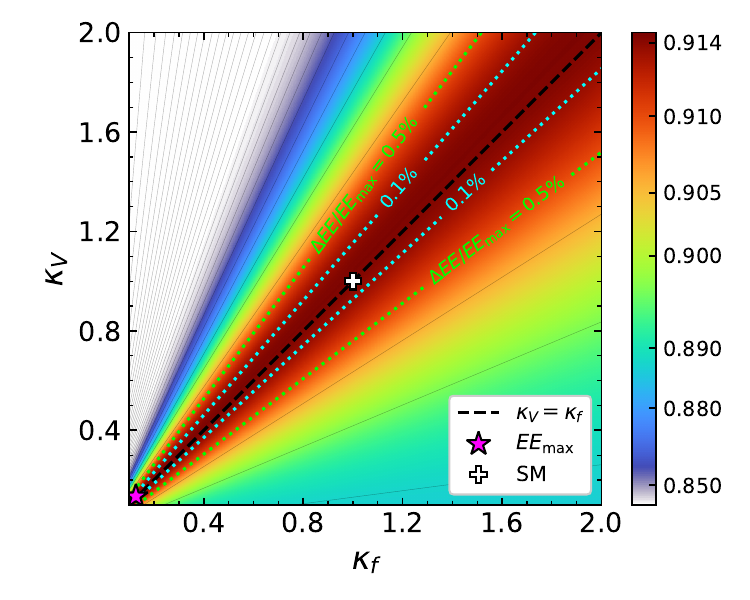}
\caption{
Distributions of the entanglement entropy using analytical formulae for the Higgs decay widths. Upper left: Dependence of the EE on the Higgs mass $m_h$. The black solid line shows results from \texttt{HDECAY}, while the orange dashed line represents the EE calculated using analytical formulae. The orange band indicates $EE_{\max}$ with its $1\sigma$ uncertainty from SM inputs and theoretical uncertainties. Other colored codes follow the same definition as in Fig.\,\ref{fig:Higgs_mass-g}. Upper right: Dependence of the EE on the $W$ boson mass $m_W$. Bottom: EE distribution in the $(\kappa_f, \kappa_V)$ plane.
}
\label{fig:higgs_mh_analytic}
\end{figure}

In the upper panels of Fig.\,\ref{fig:higgs_mh_analytic}, the $m_{h}$ and $m_{W}$ dependence of the EE is shown. 
As seen in Fig.\,\ref{fig:higgs_mh_analytic}, there is a discrepancy between the \texttt{HDECAY} results (black solid line) and those from the analytic expressions for the Higgs decay widths (orange dashed line) due to higher-order corrections~\cite{Djouadi:2018xqq}.
From the EE obtained using the analytic expressions for the Higgs decay widths, the predicted Higgs and $W$ boson masses are 
\begin{align}
m_{W} = 79.726 \pm 0.148 \,{\rm GeV} \,, \quad m_{h} = 126.54 \pm 0.28 \, {\rm GeV}. 
\end{align}
Comparing Eqs.\,\eqref{eq:mh_EE} and \eqref{eq:mw_EE}, we conclude that higher-order corrections to the Higgs decay widths are important for obtaining precise predictions for the Higgs and $W$ boson masses from the EE.
In the lower panel of Fig.\,\ref{fig:higgs_mh_analytic}, the SM point (white cross) and the contour with $\kappa_{f} = \kappa_{V}$ (black dashed line) lie in the near-maximal EE domain satisfying $\Delta EE/EE_{\rm max} \leq 0.1\%$.
This is consistent with the result shown in Fig.\,\ref{fig:2dimplot_kf_kv}.

\section{Numerical results for $EE = 1 - \sum_{i} {\rm BR}_{i}^2$ }
\label{sec:EE_1BR}

\begin{figure}[t]
\centering
\includegraphics[width=0.48\linewidth]{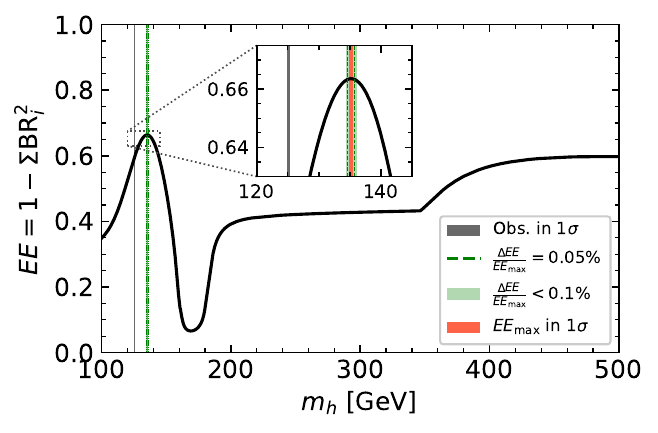}
\includegraphics[width=0.48\linewidth]{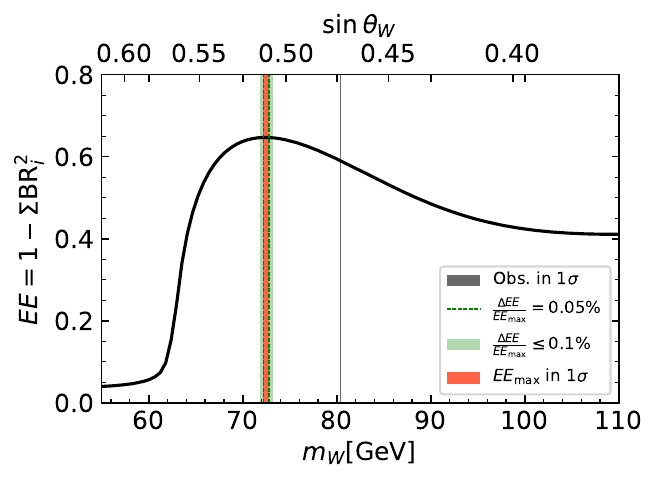}
\includegraphics[width=0.48\linewidth]{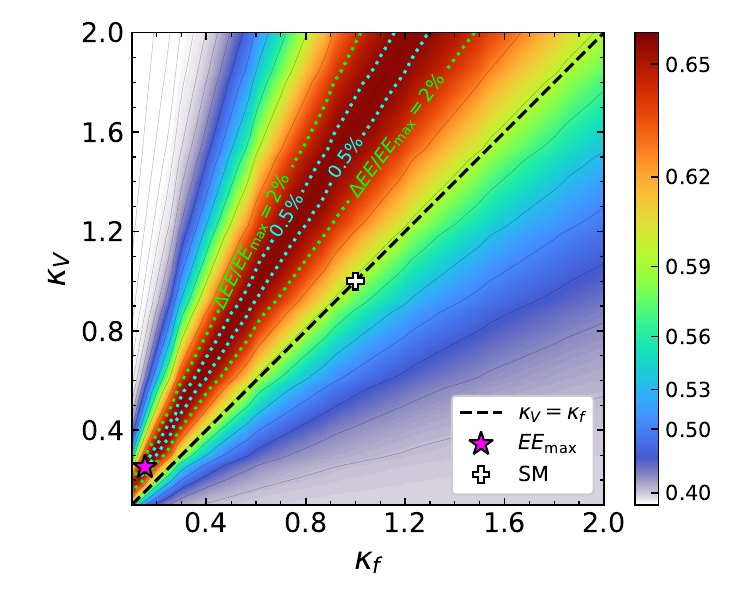}
\caption{
Entanglement entropy distribution for Standard Model parameters using $EE = 1 - \sum_{i} {\rm BR}_{i}^2$.
Upper left: EE as a function of Higgs mass $m_h$.
Upper right: EE as a function of $W$ boson mass $m_W$.
Bottom: EE distribution in the $(\kappa_{f}, \kappa_{V})$ plane.
Color coding follows the same convention as in Fig.\,\ref{fig:Higgs_mass-g}.
}
\label{fig:higgs_mh_1BR}
\end{figure}

We here show predicted SM parameters by maximizing the following simple EE
\begin{align}
EE = 1 - \sum_{i} {\rm BR}_{i}^2 \,. 
\label{eq:EE_1BR}
\end{align}
This follows from the assumption that the reduced density matrix that traced out a certain partial state for the final state is given by 
\begin{align}
\label{eq:rhof_BR}
\rho_{\rm br} = \sum_{i} {\rm BR}_{i} \ket{i} \bra{i} \,. 
\end{align}
It should be noted that the density matrix in Eq.\,\eqref{eq:rhof_BR} is of the block diagonal structure for each decay channel. 
Then, the purity is given by ${\rm tr}[\rho_{\rm br}^2] = \sum_{i} {\rm BR}_{i}^2$. 
Therefore, the linear entropy given in Eq.\,\eqref{eq:LinearEE} can be expressed by Eq.\eqref{eq:EE_1BR}.

In Fig.~\ref{fig:higgs_mh_1BR}, the parameter dependence of the EE given in Eq.\,\eqref{eq:EE_1BR} is shown. 
Our analysis yields predicted Higgs and $W$ boson masses below
\begin{align}
m_{W} = 72.497 \pm 0.313 \,{\rm GeV} \,, \quad m_{h} = 135.18 \pm 0.31 \, {\rm GeV}. 
\end{align}
Comparing Eqs.\,\eqref{eq:mh_EE} and \eqref{eq:mw_EE}, it is concluded that Eq.\,\eqref{eq:EE_1BR} does not yield plausible predictions for the Higgs and $W$ boson masses.
Furthermore, the bottom panel of Fig.~\ref{fig:higgs_mh_1BR} shows that the SM point does not lie in a region of near-maximal EE.
Based on these findings, we conclude that the EE defined by Eq.\,\eqref{eq:EE_1BR} is not applicable to the Higgs sector.

\section{Results with the Gibbs-Shannon entropy}
\label{sec:EE_Shannon}

In our analysis, we have used the entanglement entropy (EE) defined in Eq.\,\eqref{eq:EE} to calculate physical parameters such as the Higgs boson mass.
In Ref.\,\cite{Alves:2014ksa}, it was demonstrated that the maximal classical multinomial entropy (Gibbs-Shannon entropy) provides a precise prediction for the Higgs boson mass.
Here, we briefly explain their analysis and the differences from our approach.

\begin{figure}[t]
\centering
\includegraphics[width=0.48\linewidth]{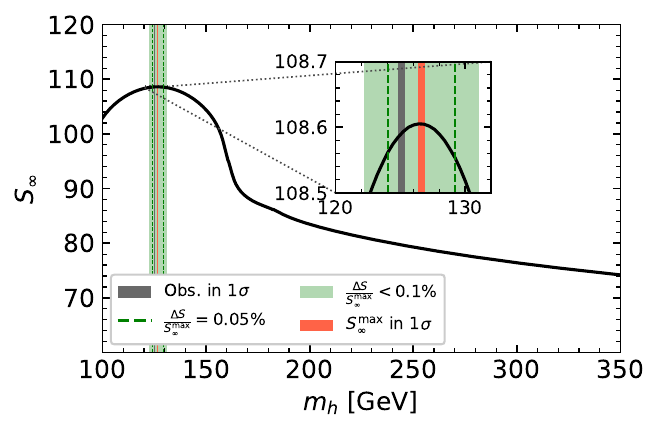}
\includegraphics[width=0.48\linewidth]{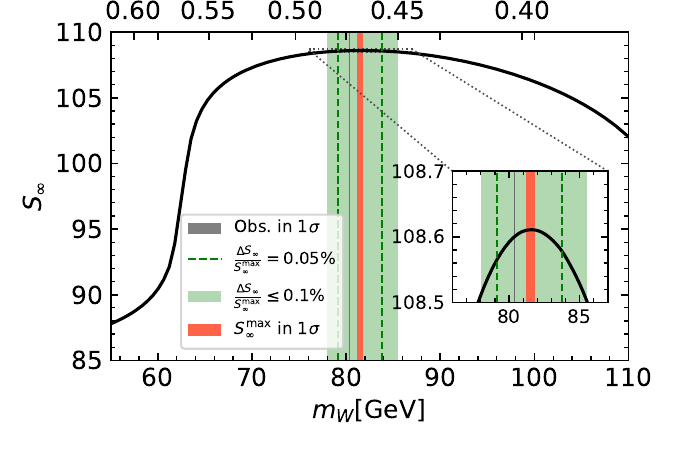}
\includegraphics[width=0.48\linewidth]{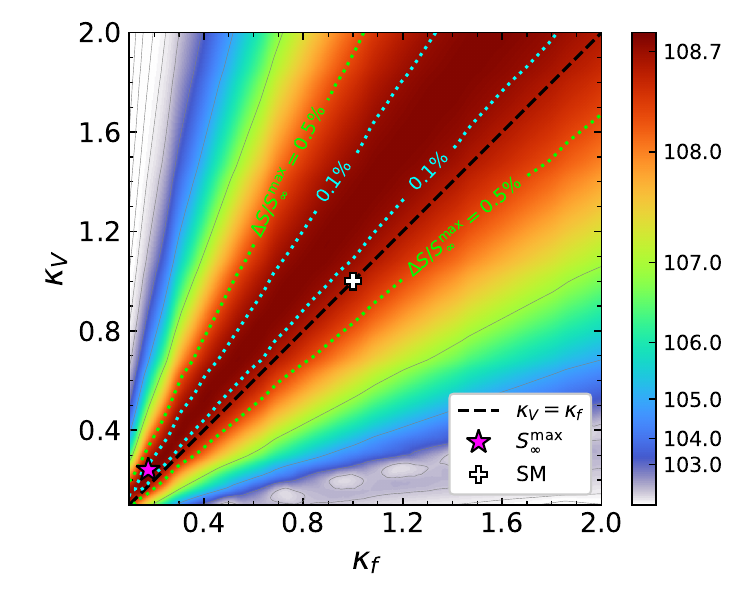}
\caption{
Gibbs-Shannon entropy (Eq.\,\eqref{eq:Shannon_Sn}) as a function of Standard Model parameters.
Upper left: Dependence on the Higgs mass $m_h$.
Upper right: Dependence on the $W$ boson mass $m_W$.
Bottom: Distribution in the $(\kappa_f, \kappa_V)$ plane.
Color coding follows the same definition as in Fig.\,\ref{fig:Higgs_mass-g}.
}
\label{fig:g2_mh_Shannon}
\end{figure}

In classical information theory, probability distributions play a fundamental role.
In Ref.~\cite{Alves:2014ksa}, the decay branching ratios of the SM Higgs boson are treated as probabilities $p_{k}$ for each channel labeled by $k$ ($k = 1, \cdots, m$), and the Gibbs-Shannon entropy is evaluated.
For $N$ Higgs bosons, the Gibbs-Shannon entropy is given by~\cite{Alves:2014ksa}
\begin{align}
S_{N} = \sum_{\left\{ n\right\}}^{N} - P(\left\{ n_{k} \right\}^{m}_{k=1}) \ln \left[  P(\left\{ n_{k} \right\}^{m}_{k=1})  \right] \,, 
\end{align}
with
\begin{align}
\sum_{\left\{ n\right\}}^{N} \equiv \sum_{n_{1} = 0}^{N} \cdots \sum_{n_{m}=0}^{N} \delta( N - \sum_{i = 1}^{m} n_{i}) \,, \quad P(\left\{ n_{k} \right\}^{m}_{k=1}) \equiv \frac{N!}{n_{1}! \cdots n_{m}!} \prod_{k = 1}^{m} (p_{k})^{n_{k}} \,, 
\end{align}
where $n_{i}$ indicates the number of Higgs bosons decaying into the channel $i$. 
The authors of Ref.~\cite{Alves:2014ksa} showed that $S_{N}$ approaches a simple expression in the large-$N$ limit (verified for $N > 10^{7}$):
\begin{align}
\label{eq:Shannon_Sn}
S_{\infty} = \frac{1}{2} \ln \left[ (2\pi N e)^{m-1} \prod_{k = 1}^{m} p_{k} \right] + O \left( \frac{1}{N} \right) \,. 
\end{align}
This implies that $S_{\infty}$ is maximized when the decay processes are democratic ($p_{1}=\cdots=p_{m}$),
a property shared by the EE in Eq.\,\eqref{eq:EE_1BR}.

The upper two panels of Fig.~\ref{fig:g2_mh_Shannon} show the dependence of the Gibbs-Shannon entropy on the Higgs and $W$ boson masses. 
Following Ref.\,\cite{Alves:2014ksa}, we use $N = 10^{9}$ in our numerical calculations.
We define the near-maximal entropy region (green domain) by $|S_{\infty}^{\max} -S_{\infty}|/S^{\max}_{\infty} \equiv \Delta S_{\infty}/S_{\infty}^{\max} \leq 0.1\%$. 
Maximizing $S_{\infty}$ yields the following predictions for $m_{h}$ and $m_{W}$:
\begin{align}
m_{h} = 126.52 \pm 0.09 \, {\rm GeV} \,, \quad m_{W} = 81.856 \pm 0.148 \,{\rm GeV} \,. 
\label{eq:mh_mW_Shannon}
\end{align}
The predicted Higgs mass deviates from the measured value by within $1.1\%$, consistent with our result in Eq.\,\eqref{eq:mh_EE}.
The predicted $W$ boson mass deviates approximately $1.9\%$ from the measured value.
The bottom panel of Fig.\,\ref{fig:g2_mh_Shannon} shows the Gibbs-Shannon entropy as a function of $\kappa_{f}$ and $\kappa_{V}$.
This figure reveals that the SM point and the contour $\kappa_{f} = \kappa_{V}$ lie outside the near-maximal entropy domain, despite the relatively accurate mass predictions shown in the upper panels.
These findings suggest that the Gibbs-Shannon entropy in Eq.\,\eqref{eq:Shannon_Sn} cannot fully characterize the Higgs sector.
Therefore, we conclude that the EE in Eq.\,\eqref{eq:EE} is more effective for determining Higgs properties and the structure of electroweak interactions.

We note that the predictions in Eq.\,\eqref{eq:mh_mW_Shannon} do not include full four-body decay channels, such as
$h\to V^{*}V^{*}\to 4f$. This omission is not because such contributions are phenomenologically irrelevant, but because their consistent incorporation into the present entropy construction is nontrivial. In our framework, the entropy is defined from a reduced density matrix obtained after specifying a Hilbert-space bipartition and tracing over the states in $\mathcal{H}_{B}$. For the two-body and single-off-shell three-body channels included in the main analysis, this prescription gives a well-defined reduced density matrix and spin factor. For full four-body final states, however, an unambiguous extension would require specifying how the observed fermions are assigned to the two intermediate vector-boson branches after the kinematic degrees of freedom are traced out. In addition, identical-fermion interference and, in the full four-fermion final state, possible $WW/ZZ$ interference effects can contribute. Therefore, the corresponding $\mathcal{P}_{i}$ would no longer be associated with a unique and straightforward channel partition in the density matrix $\rho_R$ within our present setup.

There is also a practical issue. The numerical analysis in this work follows the same computational setup as Ref.\,\cite{Alves:2014ksa}, based on \texttt{HDECAY}. While \texttt{HDECAY} provides an approximate treatment of off-shell four-body modes, a dedicated and consistent description of $h\to 4f$ decays would require a specialized tool such as \texttt{PROPHECY4F}~\cite{Denner:2019fcr}, which was not used either in our analysis or in the reference study. 

To estimate the possible size of the effect, we nevertheless considered the \textit{ad hoc} four-body off-shell contribution implemented in \texttt{HDECAY}. With this simplified prescription, we find $m_h\simeq 125.4 \,{\rm GeV}$, in agreement with Ref.\,\cite{Alves:2014ksa}. However, the same treatment gives $m_W\simeq 82.9\pm0.1 \, {\rm GeV}$ and $\kappa_V/\kappa_f\simeq 1.23$, which deviate from the measured values. We therefore interpret this exercise only as an estimate of the possible quantitative impact of four-body effects, rather than as a fully consistent incorporation of four-body final states into the entropy observable. A precision-level treatment of these effects would require a revised Hilbert-space partition and a dedicated treatment of four-fermion interference effects, which we leave for future work.

\section{The implication of EE in $Z$ boson decay \label{sec:Vboson_decay}}

\begin{figure}[t]
\centering
\includegraphics[width=0.48\linewidth]{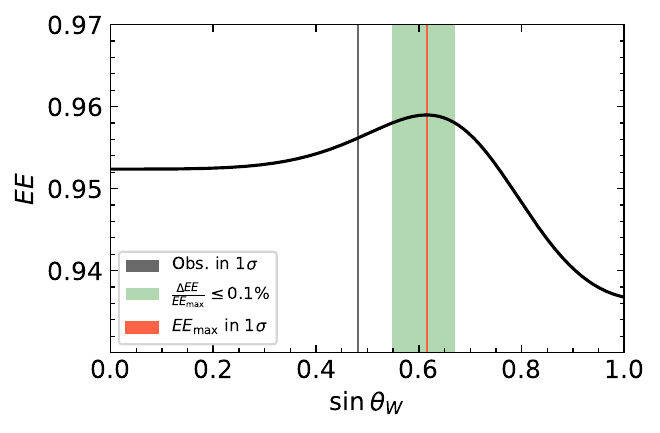}
\caption{
Dependence of the entanglement entropy on the weak mixing angle for $Z$ boson decays. Color regions follow the same convention as in Fig.\,\ref{fig:Higgs_mass-g}.
}
\label{fig:Z_boson_sW}
\end{figure}

We now apply our framework to $Z$ boson decays. The $Z$ boson decays into 11 channels: $Z\rightarrow f\bar{f} \, (f = q, \ell, \nu_{e}, \nu_{\mu}, \nu_{\tau})$ except for top quarks. 
For an unpolarized initial $Z$ boson, the leading-order decay width is given by~\cite{Donoghue:1992dd}
\begin{align}
\Gamma(Z \to f\bar{f}) = \frac{N_{c}^{f} g^2m_{Z}}{48 \pi \cos^2 \theta_{W}} \left[ c_{V}^2 + c_{A}^2 + 2 (c_{V}^2 - 2 c_{A}^2)\frac{m_{f}^2}{m_{Z}^2} \right] \sqrt{ 1 - \frac{4m_{f}^2}{m_{Z}^2} }\,, 
\end{align}
where $c_{V}$ and $c_{A}$ are the vector and axial-vector couplings in the $Zff$ vertex $-\frac{ig}{2\cos \theta_{W}} (c_{V} - c_{A} \gamma^{5})$. 
Since the EE in Eq.\,\eqref{eq:EE} depends on branching ratios, it exhibits minimal dependence on the $Z$ boson mass due to $m_{f}^2/m_{Z}^2 \ll 1$. We therefore investigate whether the EE for $Z$ boson decays can predict the weak mixing angle $\sin \theta_{W}$.

To compute the EE in Eq.\,\eqref{eq:EE}, we must determine the spin factor $\mathcal{P}_{f}$ for $Z \to f \bar{f}$. We begin by expressing the $Zff$ vertex for each fermion as
\begin{align}
-\frac{ig}{2\cos \theta_{W}} (c_{V}^{f} - c_{A}^{f} \gamma^{5}) 
= -\frac{ig}{\cos \theta_{W}} ( g_{L}^{f} P_{L} + g_{R}^{f} P_{R} ) \,, 
\end{align}
where $c_{V, A}^{f} = g_{L}^{f} \pm g_{R}^{f}$, and $P_{L, R}$ are the chiral projection operators. From this vertex, we compute the decay width $\Gamma_{ij}^{\lambda_{Z}, f}$ for $Z \to f_{i} \bar{f}_{j}$, where $\lambda_{Z}$ and $i/j$ parameterize the polarization of the $Z$ boson $(\lambda_{Z} = 0, \pm)$ and helicity for fermions ($i,j = \pm $), respectively. The spin factor is then given by
\begin{align}
\mathcal{P}_{f}^{\lambda_{Z}} = \frac{ \sum_{ij} \Gamma_{ij}^{\lambda_{Z},f} \Gamma_{ji}^{\lambda_{Z}, f} }{ \left( \sum_{i,j} \Gamma_{ij}^{\lambda_{Z}, f} \right)^2} \,, 
\end{align}
which can be expressed in terms of $g_{L/R}^{f}$ as 
\begin{align}
\mathcal{P}^{\pm}_{f}
&=
\frac{
(1-8x_f^4)\bigl[(g_L^f)^4 + (g_R^f)^4\bigr]
+ 2 x_f^2(5-12x_f^2)\,(g_L^f g_R^f)^2
- 4 x_f^2(1-4x_f^2)\,g_L^f g_R^f\bigl[(g_L^f)^2 + (g_R^f)^2\bigr]
}{
\bigl[(g_L^f)^2 + (g_R^f)^2\bigr]^2
} \,, 
\\
\mathcal{P}^{0}_{f}
&=
\frac{
(1-8x_f^4)\bigl[(g_L^f)^4 + (g_R^f)^4\bigr]
+ 16 x_f^2(1-3x_f^2)\,(g_L^f g_R^f)^2
+ 8 x_f^2(1-4x_f^2)\,g_L^f g_R^f\bigl[(g_L^f)^2 + (g_R^f)^2\bigr]
}{
\bigl[(g_L^f)^2 + (g_R^f)^2\bigr]^2
} \,, 
\end{align}
where $x_{f} = m_{f}/m_{Z}$. 
In the chiral limit ($x_f\rightarrow 0$), the spin factor is of the same form for all initial helicity: $\mathcal{P}^{\pm}_f=\mathcal{P}^{0}_f=\bigl[(g_L^f)^4+(g_R^f)^4\bigr]/\bigl[(g_L^f)^2 + (g_R^f)^2\bigr]^2$.

Fig.~\ref{fig:Z_boson_sW} shows the dependence of the EE on the weak mixing angle. The requirement of near-maximal EE yields
\begin{align}
\sin \theta_{W} = 0.615^{+0.067}_{-0.056} \,, 
\end{align}
which deviates from the measured value $\sin^2 \theta_{W}(m_Z) = 0.231$ by approximately $28\%$~\cite{ParticleDataGroup:2024cfk}. We also performed a similar analysis for $W$ boson decays, but found no meaningful results due to cancellations in the gauge coupling dependence of the EE. These findings suggest that decay processes involving global information in multiple sectors, such as Higgs boson decays, may be unique as the quantum information probes.

\twocolumngrid

\bibliography{reference}
\bibliographystyle{JHEP}

\end{document}